\documentclass[10pt,journal]{IEEEtran}
\usepackage{amsmath,amsfonts}
\usepackage{algorithmic}
\usepackage{xcolor}
\usepackage{algorithm}
\usepackage{array}
\usepackage[colorlinks=true, linkcolor=black, citecolor=black, urlcolor=black]{hyperref}
\usepackage[caption=false,font=normalsize,labelfont=sf,textfont=sf]{subfig}
\usepackage{textcomp}
\usepackage{dblfloatfix}  
\usepackage{url}
\usepackage{verbatim}
\usepackage{orcidlink}
\usepackage{graphicx}
\usepackage{cite}
\usepackage{multirow}
 
\usepackage{pifont}
\usepackage{threeparttable}
\usepackage{makecell}

\usepackage{amsmath}

\begin{document}

\title{Balancing FP8 Computation Accuracy and Efficiency on Digital CIM via Shift-Aware On-the-fly Aligned-Mantissa Bitwidth Prediction}

\author{\IEEEauthorblockN{
    Liang Zhao\orcidlink{0009-0006-1399-2995}, 
    Kunming Shao\orcidlink{0009-0002-9459-0779}, 
    Zhipeng Liao\orcidlink{0009-0003-2446-7983}, 
    Xijie Huang\orcidlink{0000-0001-5508-5433}, 
    Tim Kwang-Ting Cheng\orcidlink{0000-0002-3885-4912},~\IEEEmembership{Fellow,~IEEE,} 
    Chi-Ying Tsui\orcidlink{0000-0002-8024-2637},~\IEEEmembership{Senior Member,~IEEE,} 
    Yi Zou\orcidlink{0000-0002-4382-4670},~\IEEEmembership{Senior Member,~IEEE}
    }

\thanks{This research is supported partly by the Shenzhen SBST Research Fund No. KJZD20240903102708012, partly by NSFC Fund  No. U24B20151, as well as partly by ACCESS - AI Chip Center for Emerging Smart Systems, sponsored by InnoHK funding, Hong Kong SAR. \textit{(Corresponding authors: Kunming Shao and Yi Zou.)}}
\thanks{Liang Zhao and Kunming Shao contributed equally to this work. Liang Zhao and Yi Zou are with South China University of Technology, China (e-mail: \{202321062080, zouyi\}@mail.scut.edu.cn) and are also with MoE Engineering Research Center of DTCO of Integrated Circuits.}
\thanks{Kunming Shao, Xijie Huang, Tim Kwang-Ting Cheng and Chi-Ying Tsui are with The Hong Kong University of Science and Technology, Hong Kong SAR, China (e-mail: kshaoaa@connect.ust.hk; huangxijie1108@gmail.com; timcheng@ust.hk; eetsui@ust.hk).}
\thanks{Zhipeng Liao is with Westlake University, Hangzhou, China (e-mail: liaozhipeng@westlake.edu.cn).}
\vspace{-5mm}
}

\markboth{IEEE TRANSACTIONS ON VERY LARGE SCALE INTEGRATION (VLSI) SYSTEMS,~Vol.~12, No.~6, September~2025}%
{Shell \MakeLowercase{\textit{Zhao et al.}}: Balancing FP8 Computation Accuracy and Efficiency on Digital CIM via Shift-Aware On-the-fly Aligned-Mantissa Bitwidth Prediction}

\maketitle

\begin{abstract}
FP8 low-precision formats have gained significant adoption in Transformer inference and training. However, existing digital compute-in-memory (DCIM) architectures face challenges in supporting variable FP8 aligned-mantissa bitwidths, as unified alignment strategies and fixed-precision multiply-accumulate (MAC) units struggle to handle input data with diverse distributions. This work presents a flexible FP8 DCIM accelerator with three innovations: (1) a dynamic shift-aware bitwidth prediction (DSBP) with on-the-fly input prediction that adaptively adjusts weight (2/4/6/8b) and input (2$\sim$12b) aligned-mantissa precision; (2) a FIFO-based input alignment unit (FIAU) replacing complex barrel shifters with pointer-based control; and (3) a precision-scalable INT MAC array achieving flexible weight precision with minimal overhead. Implemented in 28nm CMOS with a 64$\times$96 CIM array, the design achieves 20.4 TFLOPS/W for fixed E5M7, demonstrating 2.8$\times$ higher FP8 efficiency than previous work while supporting all FP8 formats. Results on Llama-7b show that the DSBP achieves higher efficiency than fixed bitwidth mode at the same accuracy level on both BoolQ and Winogrande datasets, with configurable parameters enabling flexible accuracy-efficiency trade-offs.
\end{abstract}


\vspace{-2mm}
\section{Introduction}

\IEEEPARstart{T}{he} continuous advancement of large language models (LLMs) has led to a significant increase in model parameters, making low-precision quantization crucial for accelerating computation and reducing power consumption. FP8, as a low-precision floating-point format, has seen widespread adoption for Transformer inference and training~\cite{guo2025deepseek,yang2025qwen3}. Compared with integer formats, FP8 leverages its exponent bits to provide a higher dynamic range than INT8 for outliers, enabling superior accuracy~\cite{PoE,FP84DL}. However, under FP8 quantization, data with different distributions and formats (E2M5, E3M4, E4M3 and E5M2) coexist within the same network, requiring hardware support for variable aligned-mantissa bitwidths. 

\begin{figure}[t]
\centering
\includegraphics[width=0.8\columnwidth]{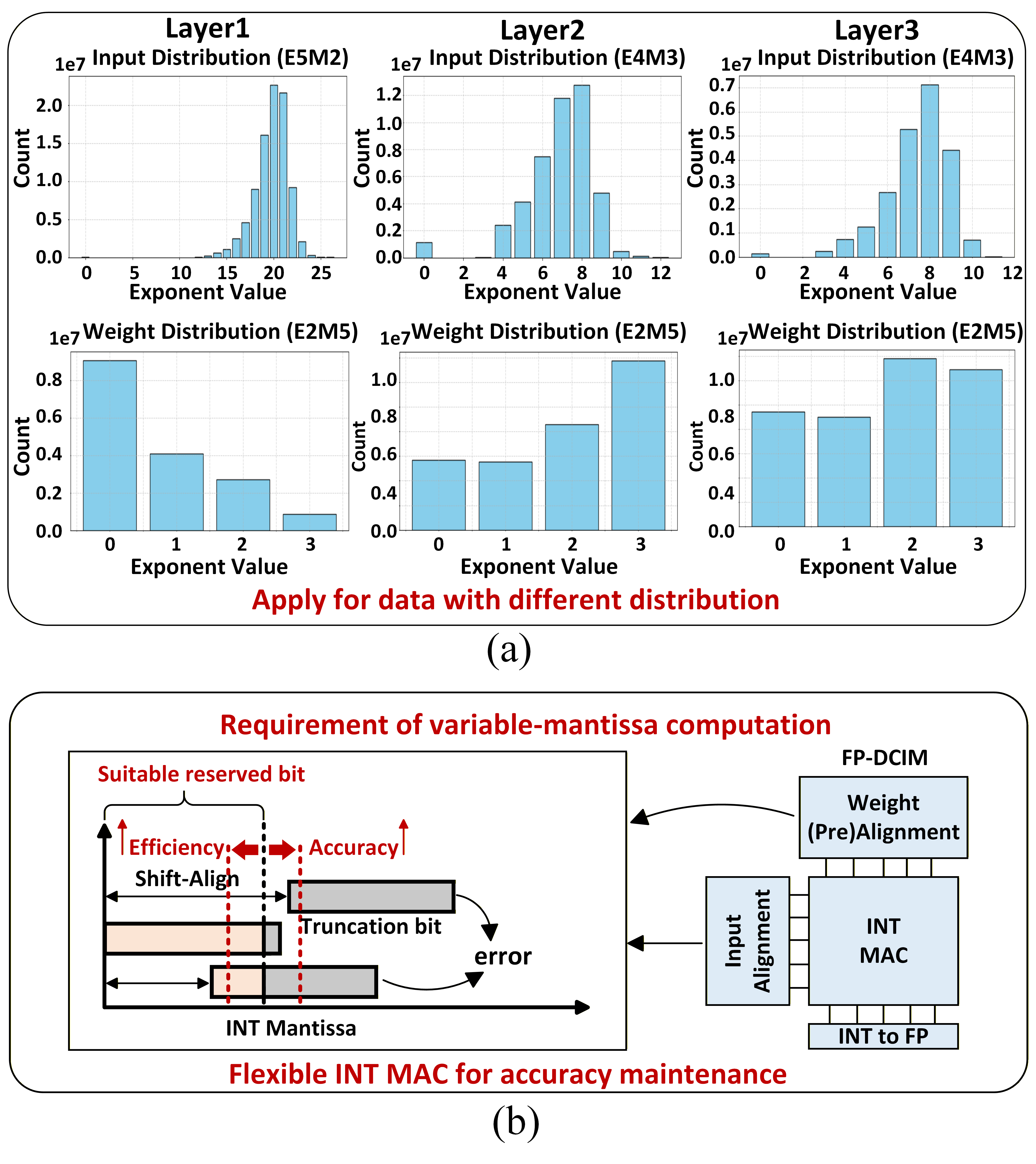}
\vspace{-5mm}
\caption{(a) FP8 parameters extracted from Llama-7b with different format. (b) Requirement of variable-mantissa computation based on FP-DCIM.}
\vspace{-2mm}
\label{FP8 applications}
\end{figure}

As shown in Fig.~\ref{FP8 applications}(a), we extracted the exponent values from three representative layers of FP8-quantized Llama-7b model under the optimal FP8 formats to analyze the data distribution. It can be observed that FP8 parameters in different formats have different exponent ranges and distributions. Furthermore, parameters of the same format extracted from different layers also exhibit differences in their distributions. 

Compute-in-memory (CIM) architectures offer lower power consumption than traditional von Neumann architectures. As shown in Fig.~\ref{FP8 applications}(b), previous FP-CIM works~\cite{tu2022redcim,diao202428nm,wen202434,guo202428,yan202428} implement floating-point MAC using INT MAC units with exponent control, where mantissas are aligned to a fixed INT format by shifting according to exponent offset before computation. For example, \cite{tu2022redcim} supports aligned-mantissa bitwidths of 8/16/24b for FP32/BF16/INT16/INT8 computations, while \cite{diao202428nm} aligns mantissas to 7/8b for INT8 and unsigned BF16 formats.

\begin{figure*}[t]
\centering
\includegraphics[width=0.8\textwidth]{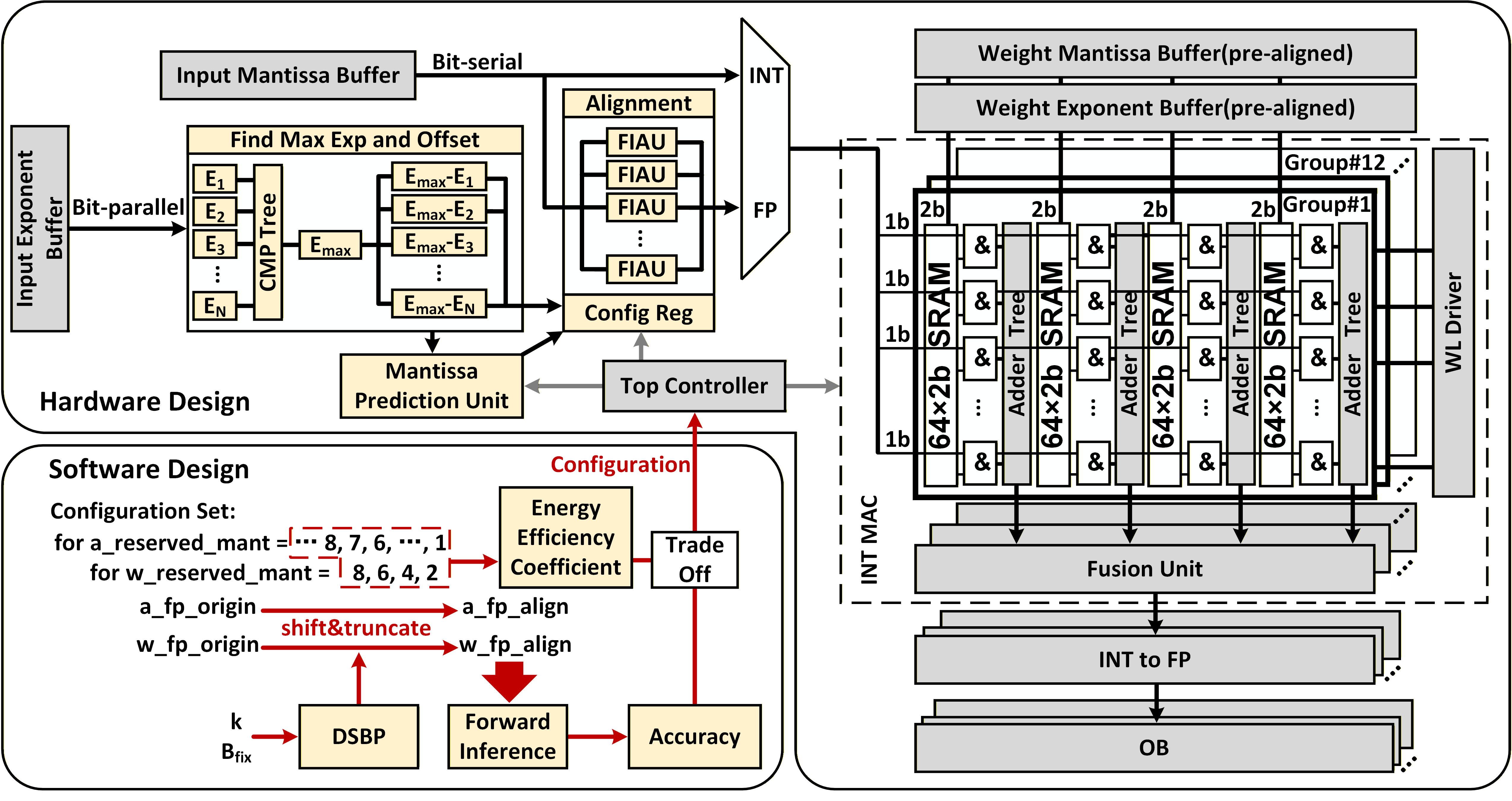}
\vspace{-3mm}
\caption{Overall framework of our software-hardware co-design Variable-Mantissa FP8 DCIM accelerator.}
\vspace{-6mm}
\label{overallarch}
\end{figure*}

However, existing designs struggle to support FP8 formats with varying aligned-mantissa bitwidths. E4M3/E5M2 formats with larger exponent ranges require more aggressive shifts to represent activation outliers, making them susceptible to truncation errors. In contrast, E2M5/E3M4 formats with smaller exponent ranges require narrower aligned bitwidths. This diversity creates a dilemma for fixed-bitwidth CIM designs:
(1) Large aligned-mantissa bitwidth~\cite{tu2022redcim,guo202428} minimizes truncation errors but degrades computational efficiency;
(2) Smaller aligned-mantissa bitwidth improves efficiency but cannot adapt to layer/channel-wise varying FP8 formats~\cite{liu2023llm}, causing substantial accuracy loss. Moreover, continuous 1$\sim$8b INT CIM arrays~\cite{tcasi1-8} incur significant precision-fusion overhead.

To address this, we propose a flexible CIM macro for FP8 inference with variable aligned-mantissa bitwidths, balancing accuracy and energy efficiency. First, dynamic shift-aware bitwidth prediction (DSBP) adaptively determines the aligned-mantissa bitwidth based on group-wise data distribution. Second, a mantissa prediction unit (MPU) enables bitwidth prediction, while a FIFO-based input alignment unit (FIAU) executes 2$\sim$12b input alignment. Third, a 2/4/6/8b INT MAC array supports flexible weight precision and efficient computation.

\vspace{-2mm}
\section{Proposed Design}

Fig.~\ref{overallarch} shows the overall hardware-software co-designed architecture. On the software side, the DSBP algorithm analyzes weight distributions offline to determine aligned-mantissa bitwidths and configures prediction hyperparameters for on-chip input bitwidth prediction. On the hardware side, the MPU predicts input aligned bitwidths on-the-fly using the configured hyperparameters. The FIAU then performs mantissa alignment based on exponent offsets and the determined bitwidths. Finally, the aligned inputs and weights are processed by a 64$\times$96 SRAM-based INT MAC array supporting 2/4/6/8b weight and 2$\sim$12b input precision through flexible fusion paths.

\vspace{-3mm}
\subsection{Dynamic Shift-aware Bitwidth Prediction (DSBP)}
Different FP8 formats exhibit varying sensitivities to a certain fixed aligned-mantissa bitwidth and yield different truncation errors. To balance accuracy and efficiency, we propose DSBP, which dynamically determines the aligned-mantissa bitwidth. The aligned bitwidth is set to 1/3/5/7b for weights and 1$\sim$11b for inputs, which is then extended by a 1b sign bit to form the aligned-mantissa for INT MAC computation on CIM. \textbf{In DSBP, a group refers to the set of weights and inputs participating in the MAC operation in each column, determining the alignment granularity.} For a 64-row SRAM array, each group contains 64 elements.

\begin{algorithm}[t]
\caption{Dynamic Shift-aware Bitwidth Prediction}
\label{alg:fp8_exp_weighted}
\begin{algorithmic}[1]
\STATE \textbf{Input parameter:} Sign ($S$), Exponent ($E$), Exponent bitwidth ($E_{\text{bit}}$), Mantissa ($M$), Group Size ($G$), Scaling Factor ($k \in \{1, 1.5, 2, 2.5, 3\}$), Fixed Bitwidth ($B_{\text{fix}}$, activations: $\{1, 2,\dots, 11\}$, weights: $\{1, 2,\dots, 7\}$)  
\STATE \textbf{Output parameter:} Aligned output tensor ($\mathbf{Y}$), Predicted aligned-mantissa bitwidth tensor ($\mathbf{B}$)
\STATE $M_{\text{bit}} = 7 - E_{\text{bit}}$
\STATE Partition $S, E, M$ into groups of size $G$
\FORALL{group $g$}
\STATE $E_{g,\max} = \max(E_g)$
\FORALL{element $i$ in group $g$}
\STATE $\text{shift}_i = E_{g,\max} - E_{g,i}$
\STATE $w_i = 2^{-\text{shift}_i}$
\ENDFOR
\STATE $B_{g,\text{dyn}} = \frac{\sum_i \text{shift}_i \cdot w_i}{\sum_i w_i}$
\STATE $B_g = \text{round\_to\_valid}(k \cdot B_{g,\text{dyn}} + B_{\text{fix}})$ \COMMENT{Weight: $\{1,3,5,7\}$, Input: $\{1,\dots,11\}$}
\FORALL{element $i$ in group $g$}
\STATE $M'_{g,i} = \text{round}\!\left(M_{g,i} \cdot 2^{B_g - M_{\text{bit}} - \text{shift}_i}\right)$
\STATE $Y_{g,i} = \text{sign}(S_{g,i}) \cdot M'_{g,i} \cdot 2^{E_{g,\max} - B_g}$
\ENDFOR
\ENDFOR
\RETURN $\mathbf{Y}$, $\mathbf{B}$
\end{algorithmic}
\end{algorithm}   

The pseudocode of DSBP is shown in Algorithm~\ref{alg:fp8_exp_weighted}. We embed this process into the MAC operation of the models. Based on the $\text{shift}_i$ values of the elements, DSBP selects appropriate values for the aligned-mantissa bitwidths to balance efficiency and truncation error. For the dynamic predict result per group $B_g$, it consists of two parts, $B_{g,\text{dyn}}$ and $B_{\text{fix}}$. $B_{g,\text{dyn}}$ is dynamically determined by the $\text{shift}_i$ values of the elements in each group. $k$ and $B_{\text{fix}}$ are hyperparameters that can be explored and adjusted offline under accuracy and efficiency requirements. $B_g$ is then rounded to the valid integer value in the search space. For weights and activations, $B_g$ is rounded to the nearest valid bitwidth of 1/3/5/7b and 1$\sim$11b, respectively.

The $\text{shift}_i$ value in $B_{g,\text{dyn}}$ reflects the contribution of each element to the MAC result. We use a weighted average based on the distribution of $\text{shift}_i$ to obtain the final $B_{g,\text{dyn}}$.
Since truncating element $i$ incurs an error proportional to $2^{-\text{shift}_i}$, setting $w_i = 2^{-\text{shift}_i}$ lets $B_{g,\text{dyn}}$ prioritize high-magnitude elements whose truncation causes the greatest accuracy loss.
For example, if all exponent values are identical, all $\text{shift}_i$ values are 0 in a group, and $B_{g,\text{dyn}}$ is 0, which means that the aligned-mantissa only needs $B_{\text{fix}}$ bitwidth without further shifting. If almost all $\text{shift}_i$ values are 3, $B_{g,\text{dyn}}$ will approach 3 to balance bitwidth and truncation error. 
If almost all $\text{shift}_i$ values are 6, $B_{g,\text{dyn}}$ still tends to stay close to 3, as their diminished contribution to the accumulation result limits their influence on $B_{g,\text{dyn}}$. $k$ fine-tunes this prediction as a hyperparameter.

\vspace{-2mm}
\subsection{Mantissa Prediction Unit (MPU) Design}
The Mantissa Prediction Unit (MPU) implements the DSBP calculation described in Algorithm~\ref{alg:fp8_exp_weighted} for on-the-fly input aligned-mantissa bitwidth prediction. Given 64 $\text{shift}_i$ values from one group, the circuit computes:
\vspace{-1mm}
\begin{equation}
B_g = k \cdot \frac{\sum_{i} \text{shift}_i \cdot 2^{-\text{shift}_i}}{\sum_{i} 2^{-\text{shift}_i}} + B_{\text{fix}}
\label{eq:shift_aware}
\vspace{-1mm}
\end{equation}

\begin{figure}[t]
\centering
\includegraphics[width=\columnwidth]{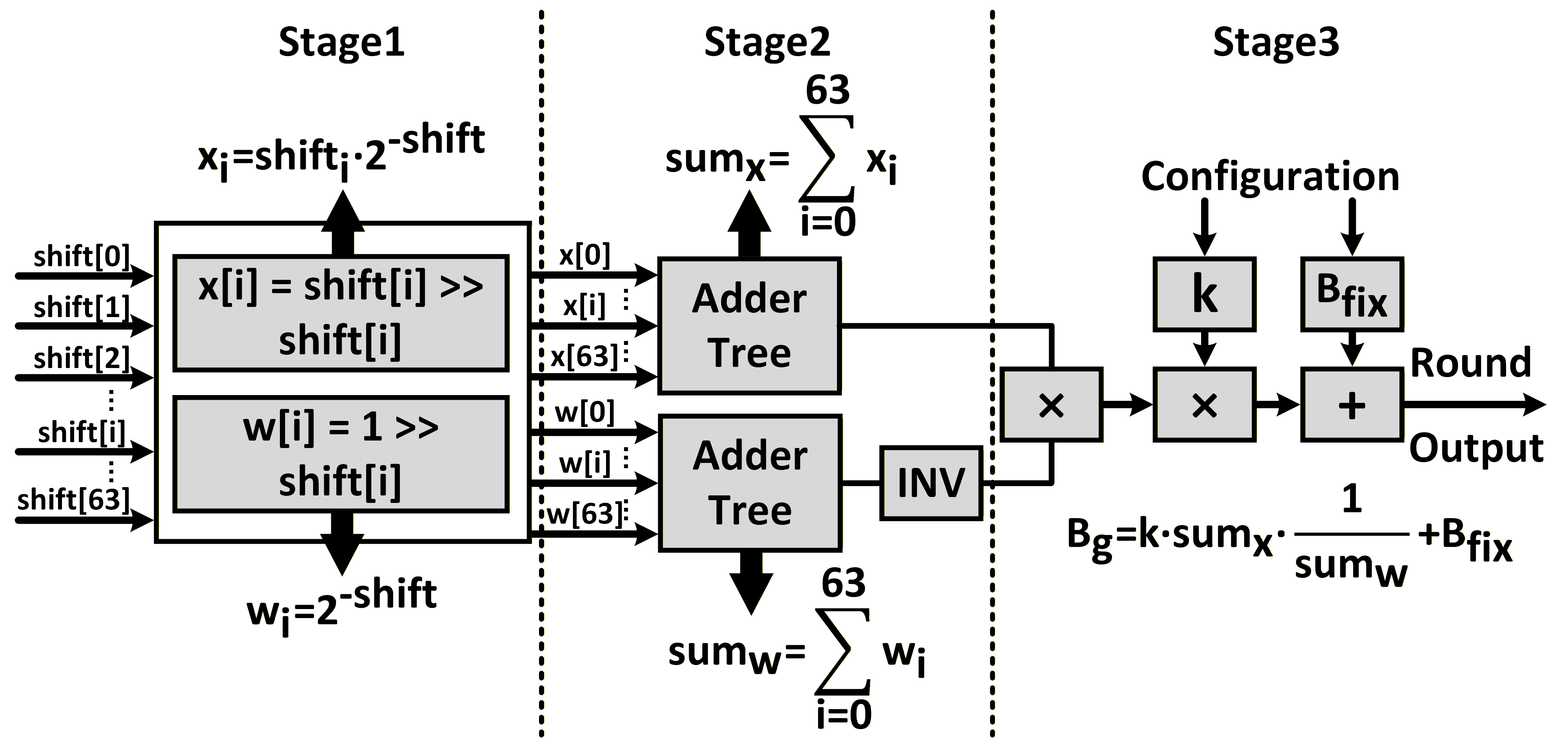}
\vspace{-7mm}
\caption{The schematic of the proposed MPU.}
\vspace{-2mm}
\label{MPU}
\end{figure}

As shown in Fig.~\ref{MPU}, the circuit employs a 3-stage pipeline architecture. In Stage 1, 64 parallel shift units compute $\text{shift}_i \cdot 2^{-\text{shift}_i}$ and $2^{-\text{shift}_i}$ by $\text{shift}_i \gg \text{shift}_i$ and $1 \gg \text{shift}_i$, respectively. Stage 2 consists of two 64-input adder trees that aggregate the shifted results. In Stage 3, division is implemented using an 8b reciprocal lookup table (LUT) to avoid expensive divider circuits. The final result is obtained by multiplying by $k$, adding $B_{\text{fix}}$, and saturating to 5b. The MPU occupies 7.0\% of the total CIM macro area and is only activated in dynamic bitwidth mode. In fixed bitwidth configurations, the MPU is clock-gated to eliminate power consumption, and $B_g$ is directly set to the predetermined fixed value. The FIAU begins alignment concurrently with MPU computation. During $B_g$ calculation, the FIAU shifts $B_{\text{fix}}$ in parallel, and once $B_g$ is returned, the FIAU determines the final bitwidth. This overlaps the MPU's 3-stage computation with FIAU operations, avoiding additional latency overhead.

\vspace{-3mm}
\subsection{FIFO-based Input Alignment Unit (FIAU)}

Previous FP-CIM works~\cite{tu2022redcim,yan202428} use parallel barrel shifters for mantissa alignment, which incurs significant area overhead and complex combinational logic depth. To alleviate this issue, we propose a FIFO-based Input Alignment Unit (FIAU) that implements alignment through pointer control.

\begin{figure}[t]
\centering
\includegraphics[width=0.7\columnwidth]{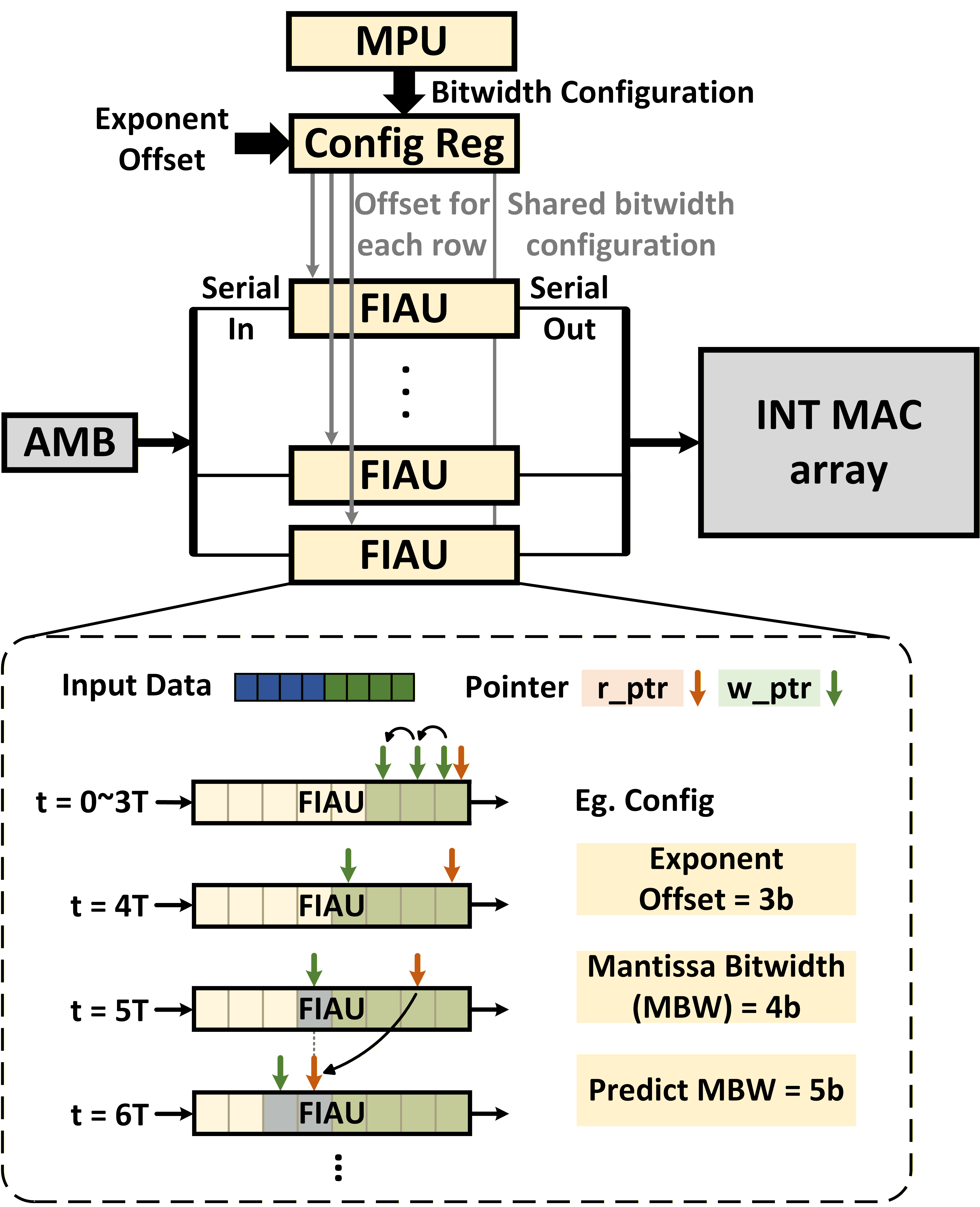}
\vspace{-3mm}
\caption{FIAU achieves alignment by controlling pointer movement.}
\vspace{-2mm}
\label{FIAU}
\end{figure}

As shown in Fig.~\ref{FIAU}, the mantissa is stored in 2's complement format and serially written from MSB to LSB, with $\text{r\_ptr}$ initially pointing to the MSB. When the read enable signal $\text{r\_en}$ is asserted, $\text{r\_ptr}$ remains at MSB for $\text{exp\_offset+1}$ cycles before advancing, implementing sign-extension equivalent to right-shift alignment. The $\text{save\_len}$ signal controls output precision: after $\text{save\_len}$ cycles, $\text{r\_ptr}$ jumps to $\text{w\_ptr}$ position for the next mantissa. This approach replaces complex barrel shifters with simple pointer logic, enabling flexible bitwidth support with minimal hardware overhead. Under the same input configuration, we measure that the 28nm synthesized FIAU achieves 21.7\% area reduction and 34.1\% power reduction compared to barrel shifters, while providing greater flexibility for variable bitwidth support.

\vspace{-3mm}
\subsection{Flexible Precision Scaling INT MAC Array}
Precision scaling architectures~\cite{tcasi1-8} accumulate column results through shift-and-add operations to obtain the final result. We have also explored a flexible precision scaling INT MAC architecture in~\cite{zhao2025flexible}, featuring reusable output fusion units with continuous 2$\sim$8b precision scaling. However, the reconfigurability of precision fusion incurs significant area and power overhead. Based on these observations, this paper adopts a dedicated CIM-based INT MAC array for FP8 computation, supporting 2$\sim$12b input and 2/4/6/8b weight. 

\begin{figure}[t]
\centering
\includegraphics[width=\columnwidth]{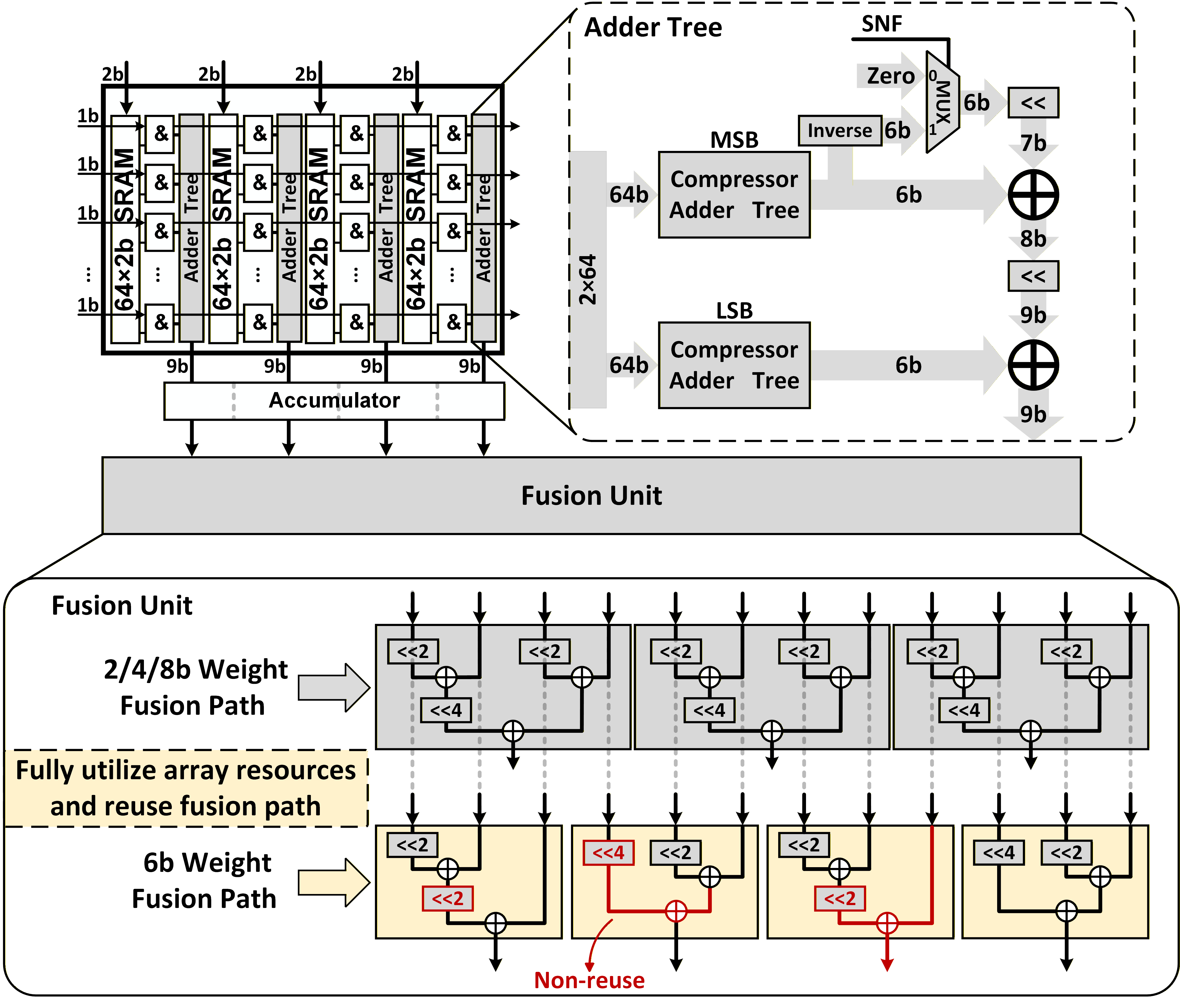}
\vspace{-8mm}
\caption{The schematic of the adder tree and fusion unit.}
\vspace{-2mm}
\label{INTMAC}
\end{figure}

As shown in Fig.~\ref{INTMAC}, our fusion unit design maximizes circuit reuse across different bitwidth modes with 64$\times$2b MAC for 2/4/6/8b weights. For 2/4/8b weight modes, scaling is achieved through the regular fusion path. However, 6b weight mode requires fusing three columns of 64$\times$2b MAC results. To fully utilize the MAC array and reuse the fusion path, we introduce the 6b weight fusion path. The red path indicates the additional circuit compared to the 2/4/8b weight fusion path, demonstrating that 6b weight path only brings a small amount of additional overhead. We use 4-2 compressors and full adders to form the adder tree for each MAC column~\cite{shao2025syndcim}. The signed number flag (SNF) indicates whether the decomposed weights are signed, following the methods in~\cite{tcasi1-8,zhao2025flexible}.

\begin{figure}[t]
\centering
\includegraphics[width=\columnwidth]{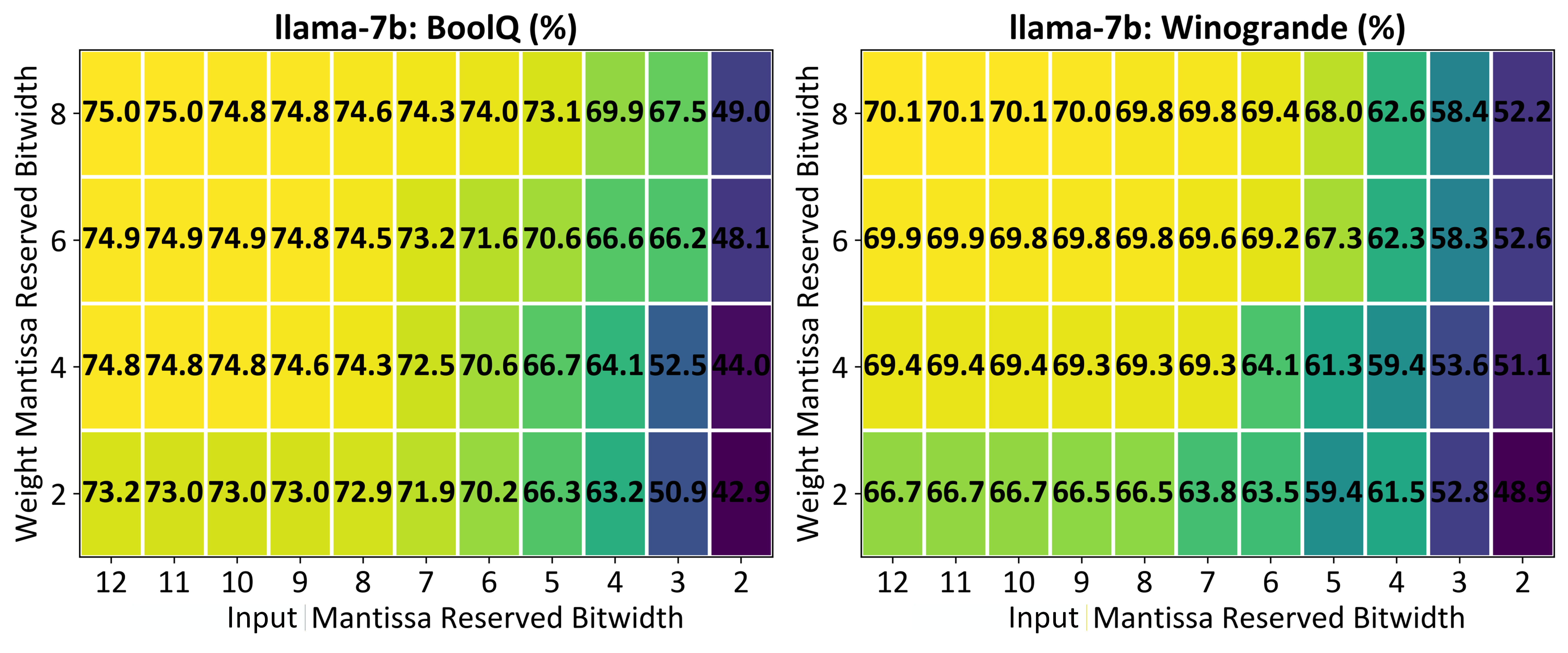}
\vspace{-8mm}
\caption{Accuracy of Llama-7b on BoolQ and Winogrande. FP8 Baseline: 75.0$\%$(BoolQ) and 70.1$\%$(Winogrande). INT8 baseline: 64.3$\%$(BoolQ) and 57.4$\%$(Winogrande).}
\vspace{-2mm}
\label{llama_accuracy}
\end{figure}

\vspace{-2mm}
\section{Validations and Evaluations}
\label{sec:experiment}

The proposed FP-DCIM is implemented in a 28nm process: SRAM cells are designed in Cadence Virtuoso, logic modules are synthesized in Synopsys Design Compiler, and placement-and-routing is performed in Cadence Innovus. Top-level integration and evaluation are completed in Virtuoso. End-to-end software experiments are conducted on NVIDIA RTX 4090 GPU, using an equivalent Python-based model for our macro.

\vspace{-3mm}
\subsection{Evaluations of Bitwidth and Accuracy}
\label{sec:accuracy}
As shown in Fig.~\ref{llama_accuracy}, we evaluate the accuracy of Llama-7b using a fixed aligned-mantissa bitwidth configuration as the baseline. Llama-7b is quantized following the method in \cite{liu2023llm}, with inputs in E4M3/E5M2 format and weights in E2M5 format. Detailed accuracy versus bitwidth exploration on the BoolQ and Winogrande datasets demonstrates that 12b input and 8b weight achieve accuracy equivalent to the FP8 baseline, establishing the upper bound of our alignment scheme.

\begin{figure}[t]
\centering
\includegraphics[width=\columnwidth]{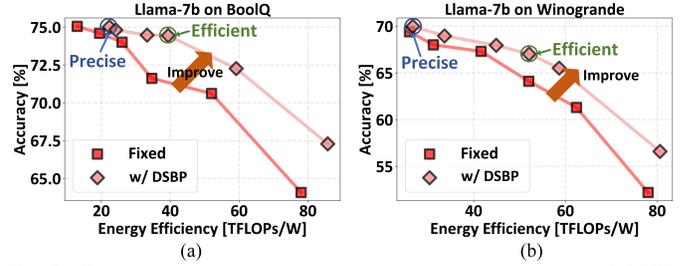}
\vspace{-9mm}
\caption{Comparison of accuracy versus energy efficiency for fixed and DSBP on Llama-7b. (a) BoolQ dataset. (b) Winogrande dataset.}
\vspace{-2mm}
\label{dynamic_strategy}
\end{figure}

Fig.~\ref{dynamic_strategy} illustrates the accuracy-efficiency trade-off between fixed and dynamic mantissa strategies on Llama-7b. \textbf{We use I/W, including sign bits, to represent the average input/weight computational bitwidths in the INT MAC array.} The macro energy efficiency is calculated as throughput divided by power, where throughput is inversely related to I$\times$W. 
As shown in Fig.~\ref{dynamic_strategy}, through dynamic prediction, DSBP effectively balances group-wise bitwidth and truncation error, forming a Pareto frontier for accuracy and efficiency. This indicates that DSBP achieves higher energy efficiency at equivalent accuracy levels compared to the baseline. Furthermore, even though DSBP and MPU introduce additional area and power overhead, both energy efficiency and throughput improve compared to fixed bitwidth mode while maintaining accuracy.

Considering the diverse requirements of practical application scenarios, we define two configuration strategies: \textbf{Precise} and \textbf{Efficient}, as marked in Fig.~\ref{dynamic_strategy}. Table~\ref{tab:efficiency} presents detailed evaluation results on the BoolQ dataset. The scaling factor $k$ and fixed bitwidth $B_{\text{fix}}$ enable flexible trade-offs between accuracy and efficiency. The \textbf{Precise} configuration ($k$=1, $B_{\text{fix}}$=6/5) prioritizes accuracy through conservative bitwidth reduction, while \textbf{Efficient} configuration ($k$=2, $B_{\text{fix}}$=4/4) aggressively reduces bitwidth for higher throughput. 
With higher $B_{\text{fix}}$=6/5, the Precise configuration has a higher base bitwidth, and yields a higher average bitwidth (BoolQ: 7.65/6.61, Winogrande: 7.02/6.02), maintaining accuracy close to the FP8 baseline. 
For the Efficient configuration, increasing $k$ to 2 and reducing $B_{\text{fix}}$=4/4 amplifies the effect of shift-aware bitwidth prediction, achieving a lower average bitwidth (BoolQ: 5.58/6.08, Winogrande: 5.18/6.02) with higher efficiency compared to Precise configuration.
This demonstrates that users can select appropriate $(k, B_{\text{fix}})$ hyperparameters based on their specific accuracy-efficiency requirements. 

To validate the generalization capability of our method across different models and datasets, we also evaluate ResNet50 on ImageNet. The Precise configuration achieves 76.1\% accuracy, matching the FP8 baseline, while the Efficient configuration achieves 74.7\% accuracy with 1.5$\times$ higher energy efficiency compared to the Precise configuration. 

\begin{table}[t]
\centering
\caption{Throughput and Energy Efficiency Comparison}      
\vspace{-3mm}     
\resizebox{0.9\columnwidth}{!}{%
\begin{tabular}{cccccc}           
\hline              
\multirow{2}{*}{\textbf{Format}} & \textbf{Avg. I/W} & \multirow{2}{*}{\textbf{$k$}} & \multirow{2}{*}{\textbf{$B_{\text{fix}}$}} & \multirow{2}{*}{\textbf{Throughput$^{(3)}$}} & \multirow{2}{*}{\textbf{Efficiency$^{(4)}$}} \\
& \textbf{(bit)} & & & &  \\ \hline
E5M3$^{(1)}$ & 4/4 & 0 & 3/3 & 0.192 TFLOPS & 77.9 TFLOPS/W \\
E5M7$^{(1)}$ & 8/8 & 0 & 7/7 & 0.048 TFLOPS & 20.4 TFLOPS/W \\
INT4$^{(1)}$ & 4/4 & - & - & 0.192 TOPS & 109.3 TOPS/W \\
INT8$^{(1)}$ & 8/8 & - & - & 0.048 TOPS & 27.3 TOPS/W \\
Precise$^{(2)}$ & 7.65/6.61 & 1 & 6/5 & 0.061 TFLOPS & 22.5 TFLOPS/W \\
Efficient$^{(2)}$ & 5.58/6.08 & 2 & 4/4 & 0.109 TFLOPS & 39.4 TFLOPS/W \\\hline    
\end{tabular}}\\[2pt]             
\raggedright\scriptsize         
$^{(1)}$50\% weight sparsity and 50\% input toggle rate. $^{(2)}$Llama-7b on BoolQ dataset. \\
$^{(3)}$Measured at 0.9V/250MHz. $^{(4)}$Measured at 0.6V/50MHz.
\vspace{-2mm}
\label{tab:efficiency}        
\end{table}

\begin{figure}[t]
\centering
\vspace{-1mm}
\includegraphics[width=\columnwidth]{figure/area_power_breakdown_combined.png}
\vspace{-9mm}
\caption{(a) Area and (b) power breakdown of the proposed CIM macro.}
\vspace{-3mm}
\label{fig:breakdown}
\end{figure}
\vspace{-2mm}
\begin{table}[t]
\centering
\caption{Comparison with SOTA FP-CIM Accelerators}
\vspace{-3mm}
\scriptsize
\resizebox{0.95\columnwidth}{!}{
\setlength{\tabcolsep}{0.5pt}%
\begin{tabular}{|c|c|c|c|c|}
\hline
 & \makecell{\textbf{CICC'24}\\ \cite{diao202428nm}$^{(1)}$ }& \makecell{\textbf{ESSCIRC'23}\\ \cite{yuan202514}$^{(1)}$} & \makecell{\textbf{ISCAS'25}\\ \cite{bazzi2025reconfigurable}$^{(2)}$} & \textbf{Ours}$^{(2)}$ \\ \hline
Process & 28nm & 28nm & 40nm & 28nm \\ \hline
Voltage & 0.55-0.9V & 0.55-1.2V & 0.7-1.2V & 0.6-0.9V \\ \hline
Freq. & 20-180MHz & 650MHz/2.4GHz & 70-435MHz & 50-250MHz \\ \hline
Area & 0.143mm$^2$ & 0.71mm$^2$ & 1.876mm$^2$ & 0.052mm$^2$ \\ \hline
SRAM & 16Kb & 4Kb & 36Kb & 6Kb \\ \hline
INT Prec. & 8b & - & 4/8b & I: 2-12b; W: 2/4/6/8 \\ \hline
FP Prec. & UBF16 & FP8(E5M2)/BF8 & FP8(E4M3) & FP8(all formats) \\ \hline
\makecell{Peak Area Eff.\\(TOPS/mm$^2$\\or TFLOPS/mm$^2$)} & \makecell{8.76(INT8)\\7.02(UBF16)} & \makecell{0.94(FP8)\\3.41(BF8)} & \makecell{0.46(INT8)\\0.31(FP8)} & \makecell{0.923(INT8)\\3.728$^{(9)}$/0.923$^{(10)}$(FP8)}\\ \hline
\makecell{Peak INT Eff.$^{(3)}$\\(TOPS/W)} & \makecell{152(INT8)$^{(4)}$\\@multiply-less NN} & - & \makecell{35.7(INT4)$^{(5)}$\\11.4(INT8)$^{(5)}$} & 27.3(INT8)$^{(5,8)}$ \\ \hline
\makecell{Peak FP Eff.$^{(3)}$\\(TFLOPS/W)} & \makecell{128(UBF16)$^{(4,10)}$\\@multiply-less NN} & \makecell{12.1(E5M2)\\66.6(BF8)} & 7.1(E4M3)$^{(5,10)}$ & \makecell{77.9(E5M3)$^{(5,8,9)}$\\20.4(E5M7)$^{(5,8,10)}$\\ 22.5(Precise)$^{(11)}$\\ 39.4(Efficient)$^{(11)}$}\\ \hline
\makecell{Dynamic\\Mantissa} & \ding{55} & \ding{55} & \ding{55} & \ding{51} \\ \hline
\makecell{CNN Acc.\\(\%)} & \makecell{ResNet50@\\ImageNet: 76.2} & - & - & \makecell{ResNet50@ImageNet:\\76.1$^{(6)}$/ 74.7$^{(7)}$} \\ \hline
\makecell{LLM Acc.\\(\%)} & \makecell{DeiT-B@\\ImageNet: 81.8} & - & - & \makecell{Llama-7b@BoolQ:\\75.0$^{(6)}$/ 74.5$^{(7)}$} \\ \hline\end{tabular}
}\\
\resizebox{\columnwidth}{!}{%
\begin{tabular}{@{}l@{}}
${(1)}$Post-silicon results. ${(2)}$Post-layout results. ${(3)}$Scaled to 28nm and measured at the lowest voltage.\\
${(4)}$50\% weight sparsity and 10\% input sparsity. ${(5)}$50\% weight sparsity and 50\% input toggle rate.\\
${(6)}$DSBP precise configuration. ${(7)}$DSBP efficient configuration. ${(8)}$Our fixed configuration.\\     
${(9)}$Aligned to 4/4b. ${(10)}$Aligned to 8/8b. ${(11)}$Measured on the Llama-7b with BoolQ dataset.\\ 
${*}$Since comparison with silicon-proven works is not fair, \cite{diao202428nm,yuan202514} are included for reference purposes. 
\end{tabular}%
  }    
\label{tab:comparison}
\vspace{-3mm}
\end{table}

\vspace{-2mm}
\subsection{Evaluations of the Proposed Hardware Architecture}
The proposed FP-DCIM architecture features a 64$\times$96 6T SRAM array, supporting variable aligned-mantissa bitwidths for weights and inputs, operating at 50$\sim$250MHz and 0.6$\sim$0.9V. Fig.~\ref{fig:breakdown} shows the area and power breakdown of our FP-DCIM macro. The input alignment unit includes FIAU, MPU, and maximum exponent processing logic. The fusion unit occupies 14.6\% of total area, with only 9.4\% for non-reused datapath, demonstrating minimal overhead. Table~\ref{tab:efficiency} presents post-layout throughput and energy efficiency. The fixed formats E5M3 and E5M7 are chosen because their I/W bitwidths (4/4b, 8/8b) correspond to common bitwidths in previous works~\cite{yuan202514, bazzi2025reconfigurable}, enabling convenient comparison. 
Based on our macro, E5M3 achieves ~4× higher efficiency than E5M7. INT8 reaches 27.3 TOPS/W, exceeding E5M7, as MPU, FIAU, and INT-to-FP circuits are disabled.  

\vspace{-2mm}
\subsection{Comparison with Previous Works}
Table~\ref{tab:comparison} compares the proposed design with state-of-the-art FP-CIM macros. \cite{diao202428nm} and \cite{yuan202514} report post-silicon measurements, while \cite{bazzi2025reconfigurable} and this work present post-layout simulation results, with energy efficiencies scaled to 28nm for fair comparison. 
This work achieves 20.4 TFLOPS/W for E5M7 (8/8b), 2.8$\times$ higher than \cite{bazzi2025reconfigurable} with 7.1 TFLOPS/W in E4M3 at the same 8/8b aligned bitwidth. 
Regarding FP8 format support, \cite{yuan202514} only supports FP8 (E5M2) and BF8, which consists of an 8b signed mantissa per value with a block-shared exponent, and \cite{bazzi2025reconfigurable} only supports E4M3 at 8/8b aligned bitwidth. In contrast, our variable aligned-mantissa architecture supports all FP8 formats from E2M5 to E5M2 at a flexible aligned bitwidth.
Furthermore, unlike previous works without dynamic bitwidth prediction, our design enables on-the-fly bitwidth adjustment based on input data distribution and offline adjustment for weights. Overall, the FIAU, compressor-based adder tree, and optimized fusion unit provide an efficient architecture for fixed FP8 computation, while DSBP with MPU further improves accuracy and efficiency.

\vspace{-2mm}
\section{Conclusion}
This paper presents a flexible FP8 DCIM accelerator supporting variable aligned-mantissa bitwidths for diverse FP8 formats. The proposed DSBP achieves an optimal balance between accuracy and efficiency. The precision-scalable INT MAC array supports 2/4/6/8b weight and 2$\sim$12b input aligned-mantissa bitwidths with minimal overhead through reusable fusion paths. Implemented in 28nm CMOS, the macro achieves 20.4 TFLOPS/W for fixed E5M7, delivering 2.8$\times$ higher FP8 efficiency than previous work. On-the-fly DSBP maintains accuracy while improving energy efficiency on Llama-7b and on ResNet50. These results demonstrate effective software-hardware co-design for efficient variable-mantissa FP8 computation in digital CIM architectures.

\vspace{-2mm}
\bibliographystyle{IEEEtran}
\bibliography{Ref}

\vfill

\end{document}